# Sub-Terahertz Channel Performance under Snowfall: From Shape-Aware Scattering to Link-Reliability Boundaries

Kefeng Huang, Jiabiao Zhao, Yuheng Song, Yapeng Ge, Jie Yang, Wanzhu Chang, Xiaoxiang Li, Wenbo Liu, Peian Li, Hong Liang, Jianjun Ma, *Member, IEEE*

***Abstract*—The terahertz (THz) band promises terabit-per-second links but is highly sensitive to snowfall. Natural snowflakes are non-spherical. Yet existing THz studies treat them as spheres under Mie theory, and no ITU-R model covers THz snow attenuation. This work combines line-of-sight measurements at 120, 140, and 160 GHz with physics-based scattering modeling. The measured loss is compared against the ITU-R P.1817-1 optical model, Mie models, and a discrete dipole approximation (DDA) for randomly oriented hexagonal-plate ice crystals, each with the Scott and Gunn-Marshall size distributions. Over the measured band, ITU-R P.1817-1 overestimates and the Mie models underestimate the loss. The shape-aware DDA-Scott model agrees best, with the lowest RMSE at every frequency. From DDA-Scott, we derive a compact modified ITU-R expression in carrier frequency and liquid-water-equivalent (LWE) rate. It reproduces the reference to within 2.5 dB/km over 100-500 GHz and 0-3 mm/h. A Rician K-factor analysis shows the channel stays LoS-dominated, so snowfall degrades the link mainly through attenuation, not multipath fading. A QPSK/16-QAM link-budget analysis then quantifies the cost of the spherical assumption. Mie-based margins overestimate the tolerable snowfall rate by ≈ 3.4 across 120-160 GHz, rising toward ≈ 5.8 in the upper transparency windows by model extrapolation. The model is further mapped into snow-limited range and adaptive-modulation switching boundaries. These results support future ITU-R recommendations for THz channels under snowfall.**

This work was supported in part by the National Natural Science Foundation of China (Grant No. 62471033), the Special Program Project for Original Basic Interdisciplinary Innovation under the Science and Technology Innovation Plan of Beijing Institute of Technology (Grant No. 2025CX11010), the Natural Science Foundation of Hebei Province (Grant No. F2026105021). First author: Kefeng Huang; Corresponding author: Jianjun Ma.

Kefeng Huang; Jiabiao Zhao; Yuheng Song; Yapeng Ge; Jie Yang; Wanzhu Chang; Xiaoxiang Li; Peian Li are with Beijing Institute of Technology, Beijing, 100081 China.

Hong Liang is with Meteorological Observation Center of China Meteorological Administration, Beijing, 100081 China

Jianjun Ma is with Beijing Institute of Technology, Beijing, 100081 China, and Tangshan Research Institute, BIT, Tangshan, Hebei, 063099 China

## I. INTRODUCTION

The terahertz (THz) frequency band (0.1-10 THz) has garnered substantial attention as a key enabler for next-generation wireless networks [1, 2]. Its ultra-wide bandwidth and short wavelengths offer the potential for terabit-per-second (Tbps) data rates, low latency, and high spatial resolution, making THz technology particularly attractive for applications such as ultra-high-speed data transmission, indoor and outdoor wireless sensing, imaging, and integrated sensing and communication (ISAC) networks [3, 4]. However, THz channel is highly susceptible to atmospheric conditions. Weather phenomena such as snow, rain, and fog introduce significant power loss, scattering, and phase distortion [5, 6], presenting critical challenges for reliable link design. Water vapor and oxygen contribute to molecular absorption, while precipitation and fog cause scattering and absorption of THz channels, potentially leading to severe power loss and degradation of bit-error-ratio [7-9].

The Radiocommunication Sector of the International Telecommunication Union (ITU-R) has established several recommendations for modeling atmospheric propagation impairments relevant to high-frequency wireless links. The ITU-R P.840 addresses attenuation due to clouds and fog [10], ITU-R P.530 provides guidance for terrestrial line-of-sight links affected by clear-air propagation and rainfall [11], and ITU-R P.676 specifies gaseous attenuation mechanisms up to 1000 GHz [12]. However, no dedicated ITU-R recommendation currently provides a physics-based or empirically validated model for snow-induced attenuation in the THz band. Although ITU-R P.1817-1 was originally developed for terrestrial free-space optical links [13], it has occasionally been adopted as a conservative reference or upper-bound approximation for estimating THz-channel attenuation under snowy conditions [14].

Snowfall is uniquely complex, owing to the heterogeneous shapes, sizes, and density variations of snowflakes [14-16]. Unlike rain droplets, which are reasonably approximated as spheres, snowflakes exhibit intricate, non-spherical structures that alter scattering behavior in ways conventional models may fail to capture [16, 17]. As shown in Table I, some typical studies of THz channel propagation under snowfall have nonetheless typically approximated snowflakes as spheres and applied Mie scattering theory to estimate channel power loss and bit error. While mathematically convenient and adequate for a baseline estimate, this approach neglects the anisotropic morphology of real snowflakes and may therefore underestimate or misrepresent shape-dependent effects, particularly at THz frequencies where the wavelength approaches the characteristic dimensions of snow crystals [18]. Such simplifications limit the accuracy of snow-channel modeling and the predictive reliability of THz channel performance under realistic conditions.

To address these limitations, this work combines outdoor sub-THz channel measurements with physics-based snowfall attenuation modeling. Measurements at 120, 140, and 160 GHz are used to quantify snow-induced power loss and validate representative ITU-R-based, Mie-scattering, and shape-aware scattering models [19, 20]. Based on the model that best agrees with the measurements, a compact modified

TABLE I
TYPICAL INVESTIGATIONS ON THZ/SUB-THZ CHANNEL PERFORMANCE UNDER SNOWFALL CONDITIONS.

| Frequency | Investigation method | Snowflake approxi. | Theory | Ref. |
|---|---|---|---|---|
| 140, 270 GHz | Theoretical & experimental | Sphere | Mie scattering | [14] |
| ≦1000 GHz | Theoretical & experimental | Sphere | Mie scattering | [16] |
| 200 GHz | Theoretical & experimental | Sphere | Mie scattering | [15] |
| ≦1000 GHz | Theoretical & experimental | Sphere | Mie scattering | [18] |
| 130-150 GHz | Theoretical & experimental | Sphere | Mie scattering | [21] |
| ≦1000 GHz | Theoretical | Sphere | Mie scattering | [22] |
| 96, 140, 225 GHz | Theoretical & experimental | - | Empirical fitting | [23] |
| 300 GHz | Experimental | Sphere | Empirical fitting | [24] |
| 77, 300 GHz | Experimental | Sphere | Empirical fitting | [25] |
| 138, 247, 340 GHz | Theoretical & experimental | - | Empirical fitting | [26] |

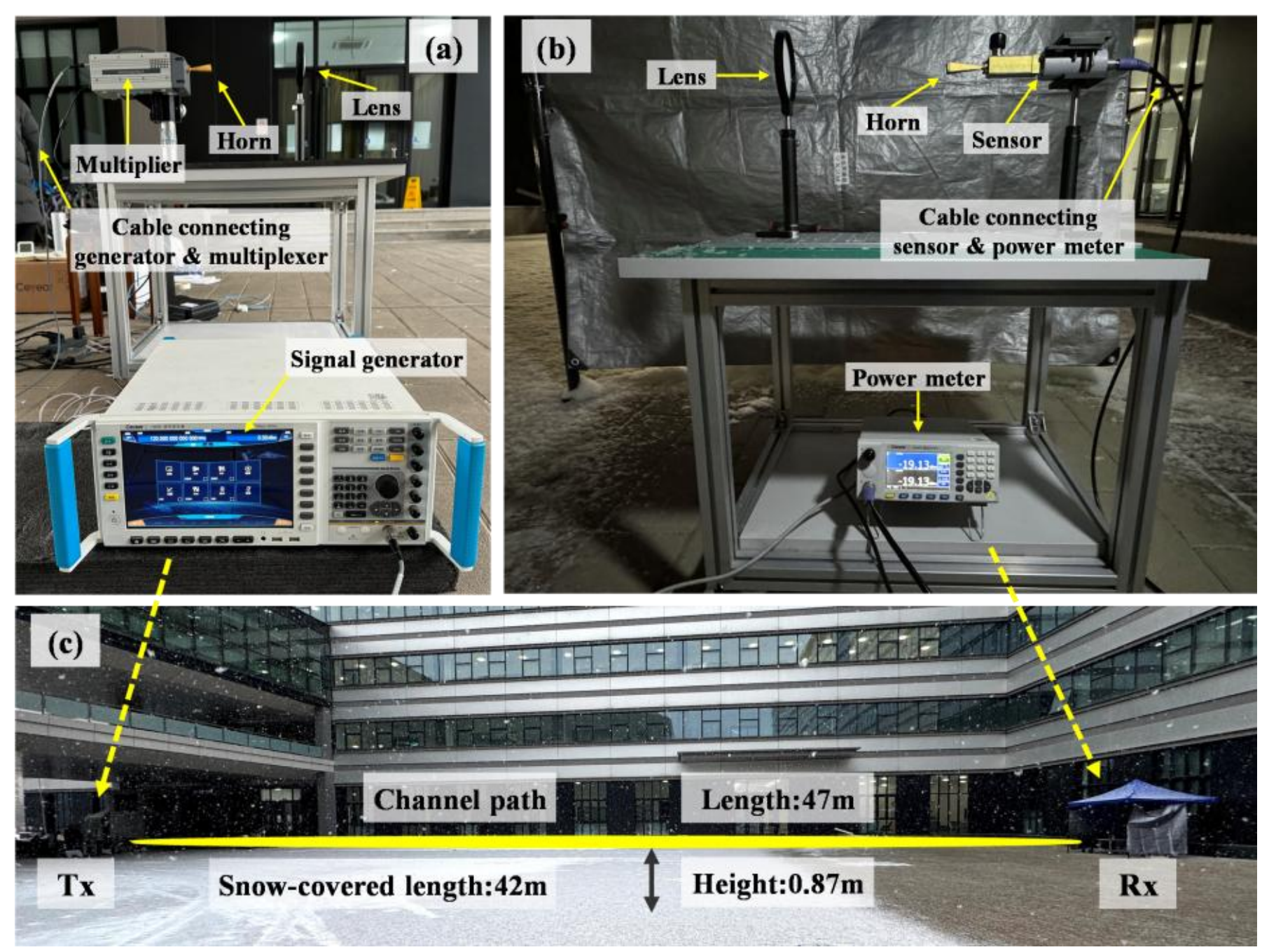

**Fig. 1.** THz channel measurement setup during the snowfall experiment at BIT. (a) Transmitter hardware. (b) Receiver hardware. (c) Outdoor channel between the Frontier Interdisciplinary Research Building at BIT Liangxiang Campus, with TX and RX safeguarded against snow.

ITU-R-type expression is developed to describe snow attenuation as a function of frequency and liquid-water-equivalent snowfall rate. The measured channel statistics are further incorporated into a Rician-fading BER analysis, linking snowfall-induced attenuation to communication performance.

## II. MEASUREMENT SETUP

The THz channel measurement was conducted at Beijing Institute of Technology, Liangxiang Campus, on January 17, 2025, during an active dry-snowfall event. The experimental channel, as shown in Fig. 1, was configured as a fixed point-to-point line-of-sight (LoS) path with a total length of 47 m, of which 42 m was directly exposed to snowfall. At the transmitter, a Ceyear 1465D vector signal generator produced the driving signal, which was up-converted to the target sub-THz frequency using a Ceyear 82406B frequency multiplier with a multiplication factor of 12. The radiated signal was launched through an HD-1400SGAH25 horn antenna integrated with a high-density polyethylene (HDPE) dielectric lens of 30 cm focal length, yielding an antenna–lens gain of 33 dBi. The receiver employed an identical horn-lens assembly, and the received power was recorded using a Ceyear 71718 power sensor. Both the transmitter and receiver were mounted 87 cm above the ground, which is well above the estimated first-Fresnel-zone radius of approximately 16 cm, thereby reducing potential ground-reflection effects. The measured beamwidth was approximately 5.7° at 140 GHz, ensuring a highly directional line-of-sight propagation path. Since the subsequent analysis relies on the relative received-power variation with respect to the clear-air reference level, the impact of absolute power-calibration uncertainty is substantially reduced. In addition, the fluctuation of the clear-air reference level was much smaller than the snow-induced attenuation observed during the experiment, confirming that the measured attenuation trend was not dominated by power-meter uncertainty.

Signal generation and power acquisition were implemented through a LAN-based instrument-control network, in which the terahertz signal generator and power meter were controlled by a host computer using LAN-SCPI commands. The protocol-layer control and data acquisition were further coordinated through custom instructions over an RS-485 bus. All measurements were continuously recorded to capture the temporal power variations introduced by snowfall.

## III. EXPERIMENTAL AND THEORETICAL ANALYSIS

Figure 2 shows the measured snow-induced attenuation at 120, 140, and 160 GHz as a function of the liquid-water-equivalent (LWE) precipitation rate. For all three carrier frequencies, the attenuation increases monotonically with LWE, indicating that a higher snow-particle concentration produces stronger extinction along the propagation path. At a fixed LWE, the measured loss also increases with frequency, which is consistent with enhanced particle-wave interaction as the wavelength decreases [16]. The scatter of the measured points around each LWE level mainly arises from the interval-averaged LWE estimate and the short-term variability of falling snow, rather than instrumental uncertainty, as discussed in Section II.

Snow-induced attenuation is governed not only by particle size, but also by snowflake shape, orientation, and dielectric properties [16]. Although Mie theory [27] provides a convenient analytical basis for estimating absorption and scattering, its spherical-particle assumption cannot represent the irregular, non-spherical morphology of natural snowflakes [17, 28]. More general electromagnetic methods can address this limitation, but with different tradeoffs. The T-matrix method is efficient for axisymmetric particles [29], whereas

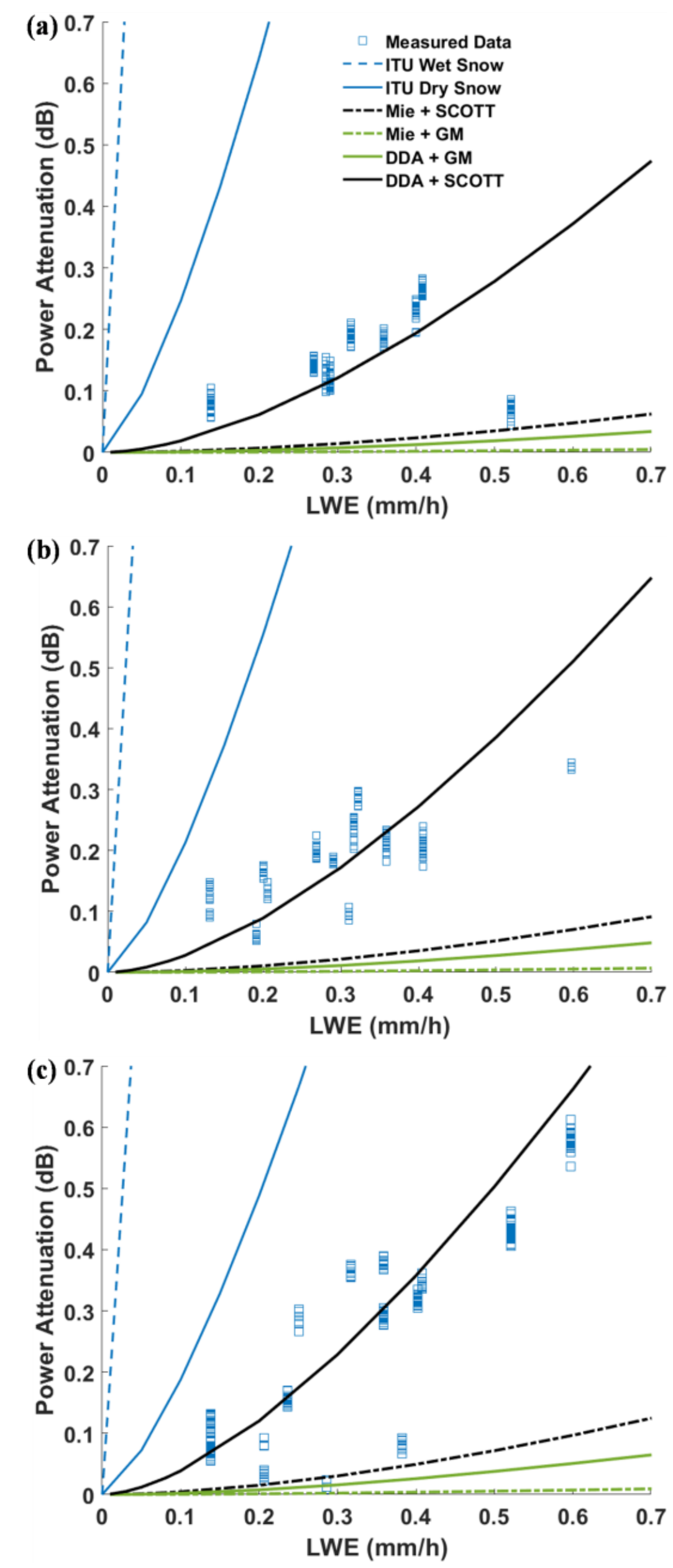

**Fig. 2.** Measured and modeled snow-induced power attenuation versus LWE at (a) 120 GHz, (b) 140 GHz, and (c) 160 GHz. (b), (c)keeps the same legend with (a).

TABLE II

PARAMETERS OF THE GM AND SCOTT PSD MODELS, R IS THE LWE PRECIPITATION RATE IN MM/H.

| PSD | $N_0$(m$^{-3}$mm$^{-1}$) | Λ(mm$^{-1}$) |
|---|---|---|
| Gunn-Marshall (GM) | $7.6 \times 10^3 R^{-0.87}$ | $5.1R^{-0.48}$ |
| Scott | $100 \times 10^3$ | $5.76R^{-0.31}$ |

full-wave differential methods such as FDTD and FEM can accommodate arbitrary geometries at substantially higher computational cost [30]. In this work, the discrete dipole approximation (DDA) is adopted because it can model particles with arbitrary shape, orientation, and composition while remaining computationally tractable for the plate-like ice crystals considered here [19, 31]. Among the tested non-spherical geometries, a randomly oriented hexagonal plate [32] provided the closest agreement with the measurements [28, 33]. A thickness-to-side-length ratio of 9:35 was used according to reported dimensions of natural plate-like snow crystals [33], and random orientation was applied to obtain an orientation-averaged extinction efficiency representative of falling snow along the propagation path.

The snowfall-induced channel attenuation is obtained by combining the extinction cross-section of individual snow particles with the particle size distribution (PSD) of the snowfall [14, 16]. Because the snow particle population varies with both space and time during a natural snowfall event [34, 35], two representative exponential PSD models are considered - the Scott distribution [14, 36] and the Gunn-Marshall (GM) distribution [35]. Other classical PSDs, such as the Marshall-Palmer [37] and Sekhon-Srivastava [38] distributions, generally yield lower attenuation than the GM model [16] under comparable conditions and are therefore not included here to avoid redundant model comparisons. The Scott and GM PSDs can be expressed in the negative-exponential form

$$N\left(r_m\right) = N_0 exp\left(-\Lambda r_m\right) \quad (1)$$

where $N_0$ is the concentration intercept parameter, Λ is the spectral slope parameter, and $r_m$ denotes the melted-equivalent particle radius. The values of $N_0$ and Λ used for the two PSDs are summarized in Table II, where $R$ is the LWE precipitation rate in mm/h.

For the DDA simulations, each snowflake is characterized by an effective radius $r_{\mathrm{eff}}$ [19], defined as the radius of a volume-equivalent sphere. Since the PSDs are formulated in terms of the melted-equivalent radius $r_m$, a radius transformation is required before integrating the DDA-derived extinction properties over the snowfall population. Assuming mass conservation during melting, $\pi r_{\mathrm{eff}}^3 \rho_s = \pi r_m^3 \rho_w$, where $\rho_s$ and $\rho_w$ are the effective snowflake density and water density, respectively. Thus, $r_m=(\rho_s/\rho_w)^{1/3} r_{\mathrm{eff}}$. In this work, $\rho_s$ was obtained by dividing the measured mass of collected snowflake samples by their total geometric volume, giving an average value of approximately 0.3 g/cm$^3$, while $\rho_w$ was taken as 1.0 g/cm$^3$. This transformation preserves particle mass and accounts for the density contrast between snow particles and their melted-water equivalents. The effective dielectric permittivity of the snow particles is then calculated using Debye-based mixing theory [16], considering the constituent volume fractions and dielectric properties of ice, air, and liquid water. For the dry-snow conditions of this campaign the liquid-water volume fraction was set to zero, reducing the three-phase formula to an ice–air mixture. The general three-phase form is retained for applicability to wet-snow cases. The resulting effective permittivity is used as the electromagnetic input for the scattering calculations [15].

The ensemble-averaged volume extinction coefficient is calculated by integrating the particle extinction cross-section over the transformed PSD [16, 39], as

$$\alpha_{\mathrm{dB/km}} = 4.343 \times 1000 \int_{r_{min}}^{r_{max}} N\left(k_{\mathrm{m}} r_{\mathrm{eff}}\right) \sigma_{\mathrm{ext}}\left(r_{\mathrm{eff}}\right) k_{\mathrm{m}} dr_{\mathrm{eff}} \quad (2)$$

where $\sigma_{ext}(r_{eff})$ is obtained from $\sigma_{ext}(r_{eff}) = Q_{ext}(r_{eff}) \pi r_{eff}^2$ and $Q_{ext}$ is the extinction efficiency computed using DDA. This formulation links the single-particle electromagnetic response to the macroscopic attenuation of the snowfall channel.

Figure 2 compares the measured attenuation with several theoretical predictions, including the ITU-R P.1817-1 snow attenuation model [13], Mie-scattering models [27] combined with the Scott and GM PSDs, and the corresponding DDA-based models [20] using the same PSDs. Although ITU-R P.1817-1 was developed for terrestrial free-space optical links rather than radio-frequency or sub-THz propagation, it is included because it provides an explicit standardized snow attenuation formulation. It therefore serves as a useful baseline for evaluating the applicability of existing ITU-R guidance to sub-THz snowfall channels. As shown in Fig. 2, ITU-R P.1817-1 substantially overestimates the measured attenuation over the tested LWE range. This discrepancy is expected because the model was derived for visible and infrared free-space optical links [13], where the scattering regime differs fundamentally from that at sub-THz frequencies. The observed overestimation motivates the recalibration of the ITU-R-type formulation for THz-band snow attenuation, as developed in the following section.

The RMSE and $R^2$ values calculated for each model are summarized in Table III. Notably, the measured sample at 120 GHz with LWE of 0.521 mm/h was found to deviate significantly from the overall trend of power attenuation due to accidental experimental errors. To ensure the reliability and representativeness of the statistical metrics, these data points was identified as an outlier and excluded from the calculation. By contrast, the Mie-Scott and Mie-GM models remain in the low-attenuation region and systematically underestimate the measured power loss, with RMSE values exceeding 0.15 dB at all three frequencies. This underestimation is mainly attributed to the spherical-particle assumption in Mie theory, which represents snowflakes as volume-equivalent spheres and therefore neglects their plate-like, dendritic, or irregular morphologies [14]. Since the electromagnetic extinction of natural snowflakes can differ substantially from that of spheres with the same equivalent radius [40], morphology-aware scattering treatment is required. The DDA-based models improve the agreement by explicitly accounting for non-spherical particle geometry. DDA-Scott achieves the lowest RMSE at each frequency (0.0415, 0.0710, and 0.1036 dB at 120, 140, and 160 GHz). Because the measured specific attenuation spans a narrow dynamic range over the 42 m path, the small total variance of the measured points inflates the relative weight of the residuals, yielding $R^2$ values that are negative for several models and, at 140 GHz, slightly negative even for DDA-Scott. We therefore treat RMSE as the primary discrimination metric, on which DDA-Scott is the closest model at every frequency, while the Mie-based models underestimate the loss by a larger and systematic margin. The difference between DDA-Scott and DDA-GM further indicates that the PSD also plays a critical role in determining the predicted attenuation, even when the same scattering method is used. Therefore, the DDA-Scott model is selected as the reference physical model for constructing the modified ITU-R formulation in the next section.

TABLE III
QUANTITATIVE EVALUATION OF DIFFERENT SNOW ATTENUATION MODELS AGAINST MEASURED DATA.

| Frequency | Model | RMSE (dB) | $R^2$ |
|---|---|---|---|
| 120 GHz | Mie-GM | 0.1696 | -6.8076 |
| | Mie-Scott | 0.1557 | -5.5818 |
| | DDA-GM | 0.1529 | -5.3423 |
| | DDA-Scott | 0.0415 | 0.5329 |
| 140 GHz | Mie-GM | 0.2035 | -9.2053 |
| | Mie-Scott | 0.1827 | -7.2210 |
| | DDA-GM | 0.1797 | -6.9518 |
| | DDA-Scott | 0.0710 | -0.2412 |
| 160 GHz | Mie-GM | 0.3253 | -2.9291 |
| | Mie-Scott | 0.2882 | -1.9374 |
| | DDA-GM | 0.2855 | -1.8827 |
| | DDA-Scott | 0.1036 | 0.6525 |

TABLE IV
POLYNOMIAL COEFFICIENTS OF THE ITU-R MODIFY MODEL.

| Function | Polynomial coefficients, highest order to constant term |
|---|---|
| $a(f)$ | $5.9810\times10^{-13}$, $-1.0992\times10^{-09}$, $8.1525\times10^{-07}$, $-3.1167\times10^{-04}$, $6.4207\times10^{-02}$, -1.3937 |
| $b_0(f)$ | $1.7739\times10^{-09}$, $7.4428\times10^{-07}$, $-2.3939\times10^{-03}$, 1.9377 |
| $b_1(f)$ | $-4.1502\times10^{-09}$, $4.3536\times10^{-06}$, $-1.3101\times10^{-03}$, $2.6806\times10^{-02}$ |

## IV. ITU-R BASED MODEL MODIFICATION

The results in Section III identify the DDA-Scott formulation as the most accurate among the evaluated models for predicting snowfall-induced attenuation in the sub-THz band. However, because DDA requires particle-level electromagnetic scattering calculations, it is not convenient for routine link-budget analysis or system-level performance evaluation. Therefore, it is necessary to develop a compact analytical surrogate that preserves the physical trend of the DDA-Scott reference while retaining the simple power-law structure commonly used in ITU-R attenuation models. The dry-snow attenuation model in ITU-R P.1817-1 [13] expresses the specific attenuation as

$$\gamma_{\mathrm{ITU}}(f,R)=aR^{b} \tag{3}$$

where $R$ is the LWE precipitation rate, and $a$ and $b$ are empirical coefficients. Although this form is simple, its fixed coefficients cannot capture the frequency-dependent attenuation behavior observed in the DDA-Scott results [40]. To improve its applicability to sub-THz snowfall channels, the attenuation model is reformulated as

$$\gamma_{\mathrm{mod}}(f,R)=a(f)R^{b(f,R)} \tag{4}$$

where $f$ is the carrier frequency in GHz. In this expression, the amplitude term $a(f)$ accounts for the frequency dependence of the extinction magnitude, while the exponent $b(f,R)=b_0(f)+b_1(f)R$ is allowed to vary with both frequency and LWE precipitation rate. This modification is essential because it enables the model to represent the nonlinear increase of attenuation with snowfall intensity, which cannot be reproduced by a constant power-law exponent. The frequency-dependent coefficients are represented by polynomial functions,

$$a(f)=\exp\left(\sum_{n=0}^{5}p_{A,n}f^{n}\right)$$
$$b_0(f)=\sum_{n=0}^{3}p_{B0,n}f^{n} \tag{5}$$
$$b_1(f)=\sum_{n=0}^{3}p_{B1,n}f^{n}$$

the polynomial coefficients were obtained by least-squares fitting to the DDA-Scott reference data over the frequency range of 100-500 GHz and the LWE range of 0-3 mm/h. The fitted coefficients are listed in Table IV in descending polynomial order. These coefficients should be used only within the fitted domain. Extrapolation outside this range is not recommended because the polynomial form is empirical and is not constrained to preserve physically meaningful behavior beyond the calibration region.

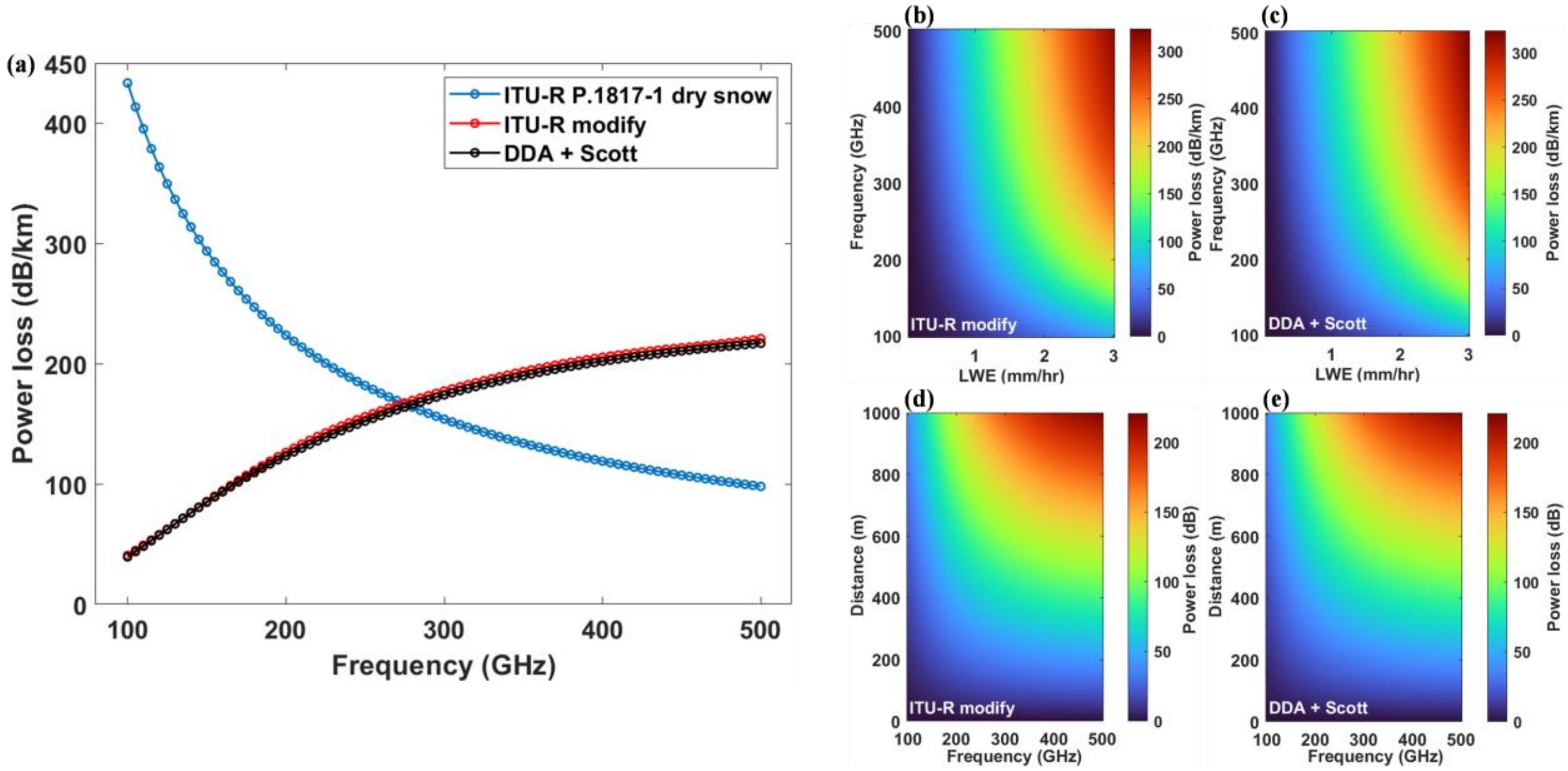

**Fig. 3.** Performance evaluation of the proposed ITU-R modify model. (a) Frequency-dependent specific attenuation comparison among the ITU-R P.1817-1 dry snow model, ITU-R modify model, and DDA-Scott at LWE = 2 mm/h. (b), (c) Specific attenuation maps of ITU-R modify model and DDA-Scott versus frequency and LWE. (d), (e) Power attenuation maps of ITU-R modify model and DDA-Scott versus frequency and propagation distance.

Figure 3 evaluates the proposed modified ITU-R-type model. Figure 3(a) compares the frequency-dependent specific attenuation at a representative LWE rate of 2 mm/h. The original ITU-R P.1817-1 dry-snow model predicts a decreasing attenuation trend with increasing frequency, which differs from the DDA-Scott reference and is inconsistent with the stronger particle extinction expected at shorter wavelengths in the sub-THz range. In contrast, the proposed model (ITU-R modify) closely follows the DDA-Scott prediction across 100-500 GHz, demonstrating that the modified parameterization effectively captures the frequency-dependent attenuation behavior.

Figures 3(b)-(e) further compare the proposed model and the DDA-Scott reference over the joint parameter space of frequency, LWE rate, and propagation distance. The specific-attenuation maps show close agreement across the fitted frequency and snowfall-intensity ranges, while the corresponding path-loss maps remain nearly indistinguishable from the DDA-Scott results. Quantitatively, the global RMSE between the proposed model and the DDA-Scott reference is 2.525 dB/km over the fitted domain, which is small relative to the attenuation range of approximately 50-250 dB/km shown in the maps. Therefore, the ITU-R modify model provides an accurate, computationally efficient, and physically consistent analytical approximation of DDA-Scott attenuation. It is consequently suitable for subsequent link-budget evaluation and BER analysis under snowy sub-THz propagation conditions.

## V. Link reliability prediction and deployment boundaries

To evaluate the communication reliability of the snowfall channel, the small-scale fading is first characterized through the Rician $K$-factor, defined as the power ratio between the dominant LoS component and the diffuse scattered components [41]. A larger $K$ indicates a more deterministic, LoS-dominated channel. For each LWE interval, the received-power samples within a stable measurement window were converted to signal amplitudes and fitted to a Rician distribution to extract $K$. As shown in Fig. 4, the fitted $K$-factor exceeds 50 dB across the measured LWE range at all three carriers, with least-squares linear fits $K$ = -8.528$R$ + 58.062, $K$ = -12.638R + 59.754, and $K$ = -6.926R + 60.452 (dB) at 120, 140, and 160 GHz, respectively. The negative slopes reflect the additional extinction and random amplitude perturbation introduced by denser snowfall [14], while the consistently high $K$ confirms a strongly LoS-dominated link, consistent with the highly directional antennas and the limited multipath produced by falling snow [14, 15, 21]. Quadratic fitting was also examined but produced stronger curvature and less reliable extrapolation over the limited measured LWE range [42]. The linear model is therefore retained. Because the power sensor integrates over a finite acquisition interval, fluctuations faster than the sampling period are averaged out, so the extracted $K$ characterizes only the resolvable fading component and should be read as an upper estimate of the channel's determinism - a point to which the robustness analysis below returns.

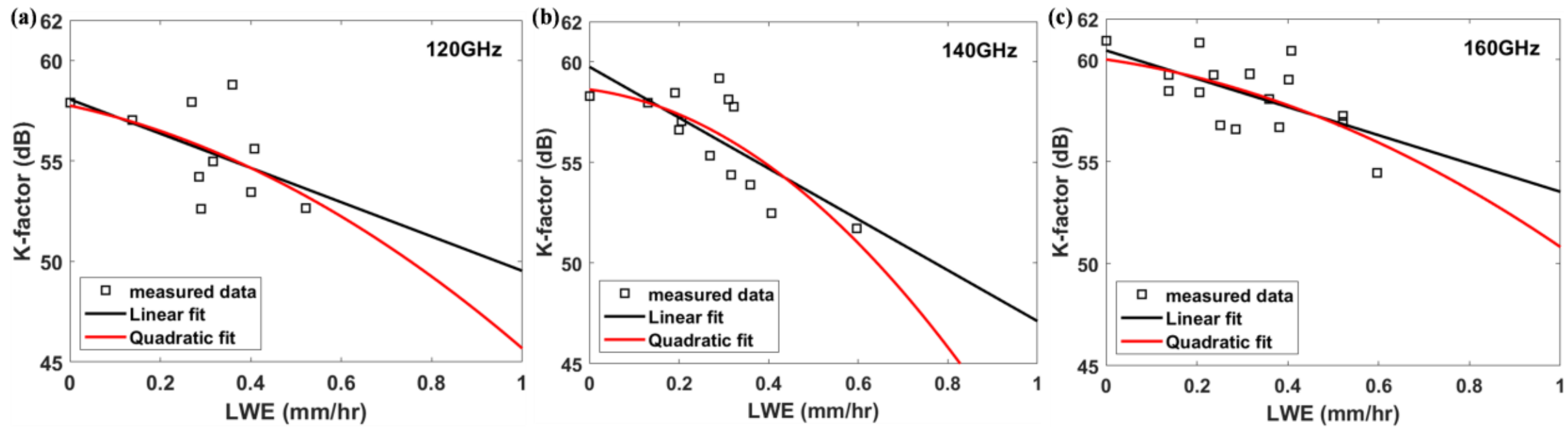

**Fig. 4.** Measured Rician K-factor and fitting curves versus LWE at (a) 120 GHz, (b) 140 GHz, and (c) 160 GHz.

Given the LoS-dominated channel, the instantaneous SNR over a path of length $d$ is governed by the deterministic link budget

$$SNR(d) = P_t + G_t + G_r - L_{FSPL}(f,d) - \gamma_{gas}(f)d - \gamma_{snow}(f,R)d - P_N \tag{6}$$

where $P_t$ is the transmit power, $G_t$ and $G_r$ the transmitter and receiver gains, $L_{\mathrm{FSPL}}$ the free-space path loss at carrier $f$, $\gamma_{\mathrm{gas}}$ the gaseous-absorption coefficient from ITU-R P.676-13 [12], $\gamma_{\mathrm{snow}}$ the snow-specific specific attenuation, and $P_N$ the receiver noise power. Reliable operation requires SNR($d$) ⩾ $\mathrm{SNR}_{\mathrm{req}}$, where $\mathrm{SNR}_{\mathrm{req}}$ is the SNR that meets the pre-FEC reliability threshold for a given format. Inverting the conditional BER expressions below at the $3.8\times10^{-3}$ hard-decision FEC limit [43] gives $\mathrm{SNR}_{\mathrm{req}}$ = 8.5 dB for QPSK and 15.2 dB for 16-QAM - a 6.7 dB separation that fixes the robustness/spectral-efficiency trade-off exploited throughout this section.

QPSK and 16-QAM are evaluated as representative formats spanning this trade-off [1, 43]. QPSK offers the higher noise immunity of the two at 2 $\mathrm{bit\cdot s^{-1}\cdot Hz^{-1}}$, whereas 16-QAM doubles the spectral efficiency to 4 $\mathrm{bit\cdot s^{-1}\cdot Hz^{-1}}$ at the cost of the 6.7 dB additional required SNR noted above. Under an AWGN channel the conditional BERs are $\mathrm{BER}_{\mathrm{QPSK}} = \frac{1}{2}erfc\left(\sqrt{\frac{\gamma}{2}}\right)$ and $\mathrm{BER}_{\mathrm{16QAM}} = \frac{3}{8}erfc\left(\sqrt{\frac{\gamma}{10}}\right)$ [44, 45], with $\gamma$ the instantaneous SNR. Since the measured channel is Rician, the average BER [46] follows from

$$\mathrm{BER} = \int_0^{\infty} BER(\gamma) f(\gamma) d\gamma \tag{7}$$

where the instantaneous-SNR PDF is

$$f(\gamma) = \frac{1+K}{\bar{\gamma}} \exp\left(-K - \frac{(1+K)\gamma}{\bar{\gamma}}\right) I_0\left(2\sqrt{\frac{K(1+K)\gamma}{\bar{\gamma}}}\right) \tag{8}$$

with $\bar{\gamma}$ the average SNR from Eq.(6), $K$ the linear Rician factor, and $I_0(\cdot)$ is the zeroth-order modified Bessel function of the first kind. Because the measured $K$-factor is generally larger than 50 dB, the Rician SNR distribution is sharply concentrated around $\bar{\gamma}$ . Therefore, the averaged BER approaches the AWGN-limited result, further indicating that snowfall mainly degrades the channel through attenuation-induced SNR reduction rather than strong fading [14].

This near-AWGN behavior is the load-bearing assumption of the section, so its sensitivity to the resolution-limited $K$-factor must be bounded. The threshold analysis was therefore repeated with $K$ swept from the measured values down to 10 dB - the conservative lower end of reported directional sub-THz LoS K-factors, against the ≈ 30 dB measured for an urban 140 GHz link in [21]. Averaging the conditional BER over Eq.(8), the additional SNR required to hold the $3.8\times10^{-3}$ threshold relative to the AWGN limit remains below ~2 dB even at $K$ = 10 dB, and is negligible (< ~0.1 dB) for K ≥ 35 dB. Propagated through Eq.(6), this contracts every range and switching boundary derived below by at most a few percent.

Fig. 5(a) presents the predicted BER versus LWE for a 1 km link. For both formats and all three carriers the BER rises monotonically with LWE as accumulated extinction lowers the received SNR, the degradation being most severe at 160 GHz where spreading loss, gaseous absorption, and snow extinction are jointly largest. At 160 GHz and the $3.8\times10^{-3}$ threshold, QPSK sustains reliable operation up to ≈ 0.55 mm/h whereas 16-QAM crosses the threshold near ≈ 0.35 mm/h, confirming that the more robust format tolerates a markedly wider precipitation range. This separation is the physical basis for an adaptive-modulation strategy [47] that falls back from 16-QAM to QPSK - and, where necessary, from higher to lower carriers - as snowfall intensity and the associated accumulated extinction increase, trading spectral efficiency for link availability. The quantitative form of this trade-off is developed next.

The model discrimination of Section III carries a direct, quantifiable cost at the link level. Under the budget Eq.(6), the maximum tolerable snowfall rate $R_{\max}(f, d)$ for a given format is the unique solution of $\gamma_{\mathrm{snow}}(f, R)\cdot d = M(f, d)$, where $M = \mathrm{SNR}_{\mathrm{clear}} - \mathrm{SNR}_{\mathrm{req}}$ is the clear-air margin. Uniqueness follows from the monotonic increase of $\gamma_{\mathrm{snow}}$ with $R$. Because the spherical Mie model systematically underestimates the specific attenuation under realistic non-spherical snowfall (Section III), it correspondingly overestimates $R_{\max}$. We therefore define the inflation factor $\eta = R_{\max,\mathrm{Mie\text{-}Scott}}/R_{\max,\mathrm{ITU\text{-}Rmodify}}$. Since the ITU-R modify model reproduces the DDA-Scott reference to within 2.525 dB/km (Section IV) and both $R_{\max}$ evaluations use the identical Scott PSD, $\eta$ isolates the particle-shape effect alone - the factor by which neglecting snowflake morphology inflates the apparent snowfall tolerance. Moreover, for a power-law $\gamma_{\mathrm{snow}} = cR^b$ one has $R_{\max}$

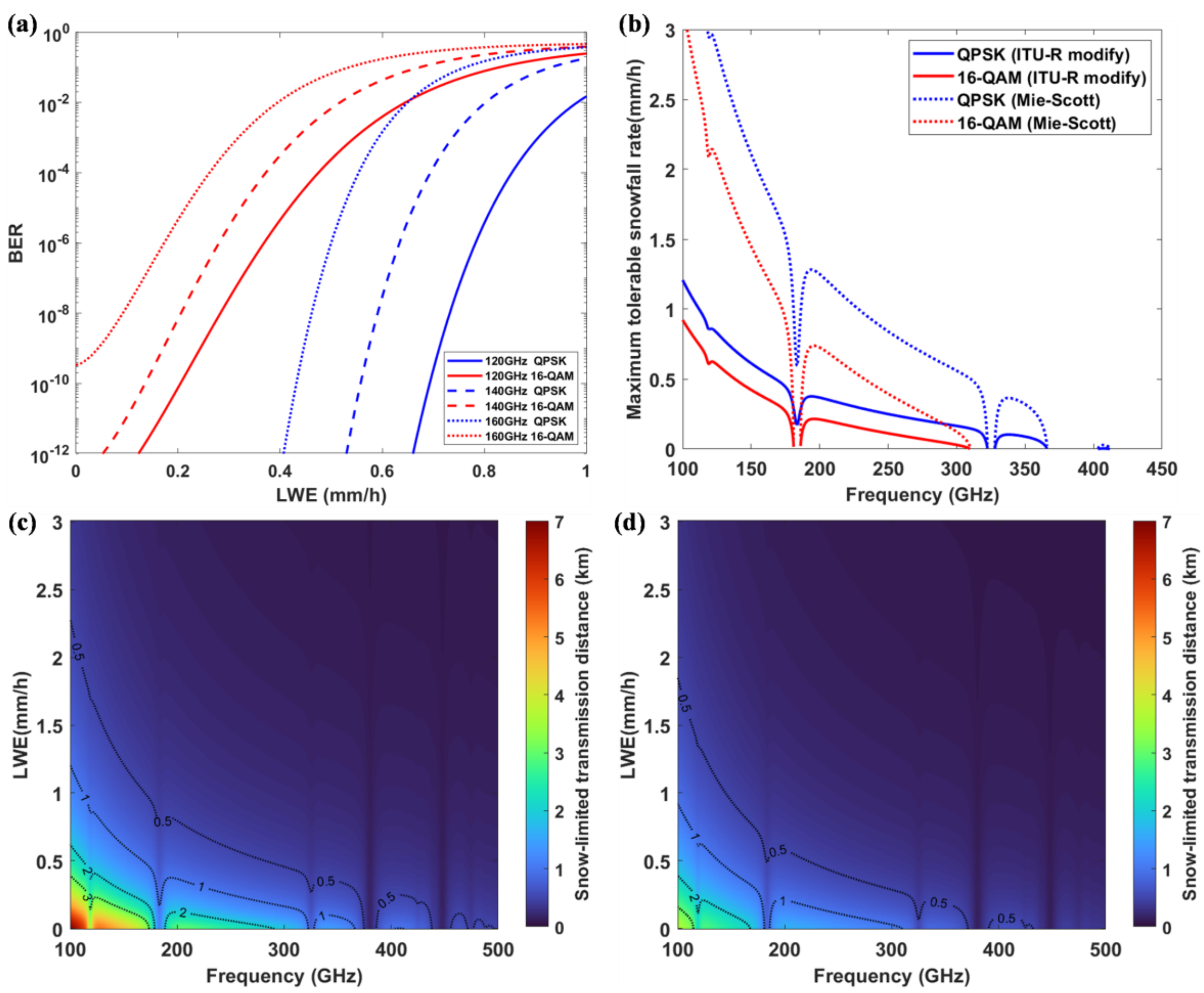

**Fig. 5** (a) BER performance of QPSK and 16-QAM based on the ITU-R modify model versus LWE at 120, 140, and 160 GHz. Channel distance 1 km, temperature 0°C, relative humidity RH 50%, transmitted power 0 dBm, transmitter and receiver gains are both 30 dB, receiver noise power -60 dBm. (b) $R_{\max}$ across 100-450 GHz under identical link budget and hardware configurations adopted in (a), contrasting the ITU-R modify model against the Mie-Scott model. (c), (d) Snow-limited transmission distance $d_{\max}$ for (c) QPSK and (d) 16-QAM modulation formats versus carrier frequency and LWE.

$= (M/(c\cdot d))^{(1/b)}$, so $\eta = (c_{\text{modify}}/c_{\text{Mie}})^{(1/b)}$ is independent of the clear-air margin and hence of the modulation format; the residual difference between the QPSK and 16-QAM values reported below (3.43 vs 3.44) arises only from the weak LWE-dependence of the exponent $b(f, R)$.

Fig. 5(b) compares $R_{\max}$ across 100-450 GHz for a 1 km link under both attenuation inputs. The sharp dips near 183 and 325 GHz coincide with the water-vapor absorption lines, where the clear-air margin collapses and $R_{\max} \to 0$ for both models; η is undefined in their immediate vicinity and is reported only within the intervening transparency windows. Beyond ≈ 410 GHz the 1 km clear-air margin is exhausted by spreading and gaseous loss alone, so $R_{\max} \to 0$ irrespective of the snow model, which sets the upper limit of the plotted range. Within the transparency windows, η increases from ≈ 3.36 at the low-frequency edge to ≈ 5.78 toward the upper band, the growth reflecting the increasing shape-dependent extinction contrast as the wavelength approaches the crystal dimension. Across the narrow 120–160 GHz window $\eta$ is nearly constant at ≈ 3.43 (QPSK) and ≈ 3.44 (16-QAM). In engineering terms, a sub-THz link dimensioned with a spherical-model snow margin will violate its availability target during snowfall that the design model classifies as benign - by a factor of roughly three at the measured carriers, rising toward six at the upper band. This is a systematic, non-conservative design error, not a second-order refinement [15],[17].

The closed-form structure of the ITU-R modify model also permits the snow-limited transmission distance $d_{\max}(f, R)$ to be evaluated continuously as the largest d satisfying $\text{SNR}(d) = \text{SNR}_{\text{req}}$ in Eq.(6), in which the spreading, gaseous, and snow terms all increase with $d$. Figs. 5(c) and 5(d) map $d_{\max}$ for QPSK and 16-QAM over 100-500 GHz and 0-3 mm/h. In the lower-frequency window (100-150 GHz) under light snowfall ($R < 0.5$ mm/h), QPSK sustains $d_{\max}$ of roughly 2-7 km. For the spectral-efficiency-optimized 16-QAM the reachable range contracts to ≈ 1-4 km. For both formats, jointly increasing the carrier frequency and snowfall rate raises molecular absorption and ice-particle extinction together, confining $d_{\max}$ below 1 km across most of the upper band - a quantitative statement of the range-throughput-weather trade-off that governs cold-climate sub-THz deployment.

For a fixed deployment distance $d$, the locus on which $d_{\max}$ for 16-QAM equals $d$ defines the modulation-switching boundary $R_{sw}(f, d)$. For $R_{sw}$ the link must fall back from 16-QAM to QPSK [47]. The corresponding QPSK locus $R_{out}(f, R)$, beyond which even QPSK fails to close, bounds the snow-induced outage region. Unlike fixed-distance BER curves, these boundaries furnish deployment-oriented design rules [11, 14]. For a given carrier and distance, the modulation-fallback duty cycle and the outage probability follow directly from the local snowfall-rate exceedance statistics as $P(R > R_{sw})$ and $P(R > R_{out})$, respectively. As an illustration, the winter-season snowfall climatology of the Beijing region is dominated by light snow, with on the order of 11-14 snow days and a December–February liquid-water-equivalent accumulation of ≈ 8 mm [48, 49], so the LWE rate exceeds the 16-QAM and QPSK switching thresholds (0.35 and 0.55 mm/h for the representative 1 km, 160 GHz link) for only a small fraction of the season. Evaluating $P(R > R_{sw})$ and $P(R > R_{out})$ from the China Meteorological Administration (CMA) winter LWE statistics for Beijing [48], the high-throughput 16-QAM format is sustainable for ≈ 99.6% of the winter season (snow-limited for only ~8 h), while the more robust QPSK keeps the link available for ≈ 99.8% (~4 h snow-limited). The adaptive fallback thus recovers roughly half of the residual snow-induced unavailability.

## VI. Conclusion

Reliable cold-climate sub-THz design needs accurate snow-attenuation models, but snowflake morphology has been largely ignored in THz studies. We combined measurements at 120, 140, and 160 GHz with physics-based scattering modeling. The measured loss rose monotonically with both snowfall rate and frequency. ITU-R P.1817-1 overestimated the loss, and the Mie models underestimated it. The shape-aware DDA-Scott model gave the closest agreement, with the lowest RMSE at every frequency (0.042–0.104 dB). Using DDA-Scott as reference, we derived a compact modified ITU-R expression. It reproduces the reference to within 2.5 dB/km over the fitted 100-500 GHz and 0-3 mm/h domain, giving an efficient surrogate for link-budget analysis in that range.

The measured Rician $K$-factor indicates a strongly LoS-dominated channel, where snowfall acts mainly through attenuation rather than fading. A sensitivity sweep confirms the BER conclusions hold even at $K = 10$ dB. The link-layer analysis quantified the penalty of the spherical assumption. Across 120-160 GHz, Mie-based margins overestimate the tolerable snowfall rate by ≈ 3.4. This is a systematic, non-conservative design error, not a minor refinement. Model extrapolation suggests the factor grows toward ≈ 5.8 in the upper transparency windows, but that band is unmeasured and awaits validation. We also mapped snow-limited range and adaptive-modulation switching boundaries. For Beijing's light-snow winters, the LWE rate exceeds the switching thresholds only briefly. The modeled availability is therefore high (≈ 99.8% for QPSK and ≈ 99.6% for 16-QAM on a 1 km, 160 GHz link). These figures mainly reflect the region's low snowfall frequency, with adaptive fallback recovering about half of the small residual unavailability. They derive from seasonal climatology and should be refined with hourly LWE exceedance data. Future work should span multi-year campaigns, more crystal habits, and higher frequencies to validate and generalize the model.